\documentclass{optica-article}

\journal{opticajournal} 

\articletype{Research Article}

\begin{document}

\title{Spectrally engineered collinear type-0 SPDC source with enhanced spectral brightness for entanglement distribution}

\author{Dong-Gil Im,\authormark{*} Jungmo Lee, Kyungdeuk Park, Dongkyu Kim, Yonggi Jo, and Yong Sup Ihn\authormark{$\dag$}}

\address{Ageny for Defense Development, Daejeon 34186, Korea}

\email{\authormark{*}eastgil@add.re.kr} 
\email{\authormark{$\dag$}yong0862@add.re.kr} 

\begin{abstract*} 
Entangled photon sources with high spectral brightness are important resources for photonic quantum information processing, particularly in quantum communication and quantum networking where usable photon flux of entangled photons is often constrained by channel loss and source inefficiency. 
 Here, we demonstrate a spectrally engineered type-0 spontaneous parametric down-conversion (SPDC) source with enhanced spectral brightness for entanglement distribution. By pumping a 30-mm ppKTP crystal with an ultra-narrowband laser slightly detuned from degeneracy, photon-pair generation is concentrated into a narrow spectral bandwidth while retaining the strong nonlinear interaction of type-0 phase matching. The source produces a coincidence rate of 44.6 kHz corresponding to a detected spectral brightness of 0.507 MHz mW$^{-1}$nm$^{-1}$. We further integrate the source into a Sagnac interferometer to generate polarization-entangled photon pairs and demonstrate entanglement distribution through a 2.56 km free-space round-trip channel. Our results show that spectral engineering provides a practical route to compact, spectrally bright entangled-photon sources for quantum communication applications.
\end{abstract*}

\section{Introduction}
Entangled photon pairs generated via spontaneous parametric down-conversion (SPDC) constitute a central resource for quantum information processing, enabling protocols in quantum communication \cite{Gisin07,Ursin07}, distributed quantum computing \cite{Main25}, and quantum networking \cite{Liao18,Li22,Lai25}. In these applications, the useful performance of a photon-pair source is not determined solely by the total number of generated photons, but by the photon-pair flux available within a usable spectral bandwidth. High spectral brightness is particularly important in practical quantum communication systems, where spectral filtering, background suppression, and coupling to optical components can strongly affect the coincidence and entanglement distribution rates.

Depending on the phase-matching and pump conditions, SPDC sources can be engineered to optimize brightness \cite{Steinlechner12,Lee25}, spectral properties \cite{Burlakov01,Baek08,Lee15}, and collection efficiency \cite{Ursin14,Schwaller22}, making the choice of SPDC configuration a critical factor for practical entangled-photon sources. Among these configurations, type-0 SPDC is attractive since it provides a large effective nonlinear interaction \cite{Jabir17}, enabling efficient photon-pair generation. However, conventional type-0 SPDC source often produce braodband emission, and the usable spectral brightness can be limited when only a narrow portion of the generated spectrum is relevant for a given quantum information task. Spectral filtering can define such a bandwidth \cite{Ursin18,Ursin20,JHK21}, but it discards photon pairs generated outside the filter window and therefore reduces the usable photon flux. In addition, collinear geometries offer an additional advantage by enabling simple spatial mode matching and efficient collection into a single-mode fiber \cite{Sarker04,Park25}, which are desirable for compact and deployable quantum light sources. Yet, in collinear type-0 SPDC, the generated photons share the same spatial mode and polarization and can strongly overlap spectrally, making deterministic photon-pair separation difficult without spectral filtering. Consequently, type-0 sources that require separated photon pairs have typically relied on noncollinear geometries \cite{Moon19}.

Spectral engineering provides a way to address these limitations at the generation stage, rather than relying on post-generation filtering or a change in emission geometry \cite{Wong13,Pan18}. In SPDC sources, such engineering has commonly been used to control emission bandwidth, frequency correlations, and spectral purity through crystal design, phase matching, or group-velocity matching \cite{Carrasco06,Mosley08}. A less explored route is to reshape the joint spectral structure by detuning an ultra-narrowband pump from the degenerate phase-matching condition. This approach can concentrate type-0 emission into a narrow usable bandwidth while preserving the advantages of collinear operation.

In this work, we demonstrate a spectrally engineered collinear type-0 SPDC source with enhanced detected spectral brightness for entanglement distribution. A 30-mm-long ppKTP crystal is pumped by an ultra-narrowband continuous-wave laser slightly detuned from the degenerate phase-matching condition, producing a spectrally split photon-pair spectrum. With a coincidence rate of 44.6 kHz, the source achieves a detected spectral brightness of 0.507 MHz mW$^{-1}$nm$^{-1}$, without correcting for optical transmission or detector efficiencies. The spectrally engineered emission also enables practical separation of the generated photons into distinct optical paths using a single dichroic element without loss, allowing the source to be incorporated into a Sagnac interferometer for polarization entanglement generation.
We further demonstrate the utility of the source by distributing polarization entanglement through a 2.56 km free-space round-trip channel. The preservation of photon-pair correlations and polarization entanglement after free-space transmission shows that the spectrally engineered type-0 source is compatible with practical entanglement distribution architecture. Our results establish spectral engineering as a useful strategy for realizing compact, spectrally bright entangled-photon sources for quantum communication and quantum networking applications.

\section{Results and Discussion}

\subsection{Photon pair generation}
\begin{figure}[t!]
                \centering
                \includegraphics[width=0.53\textwidth]{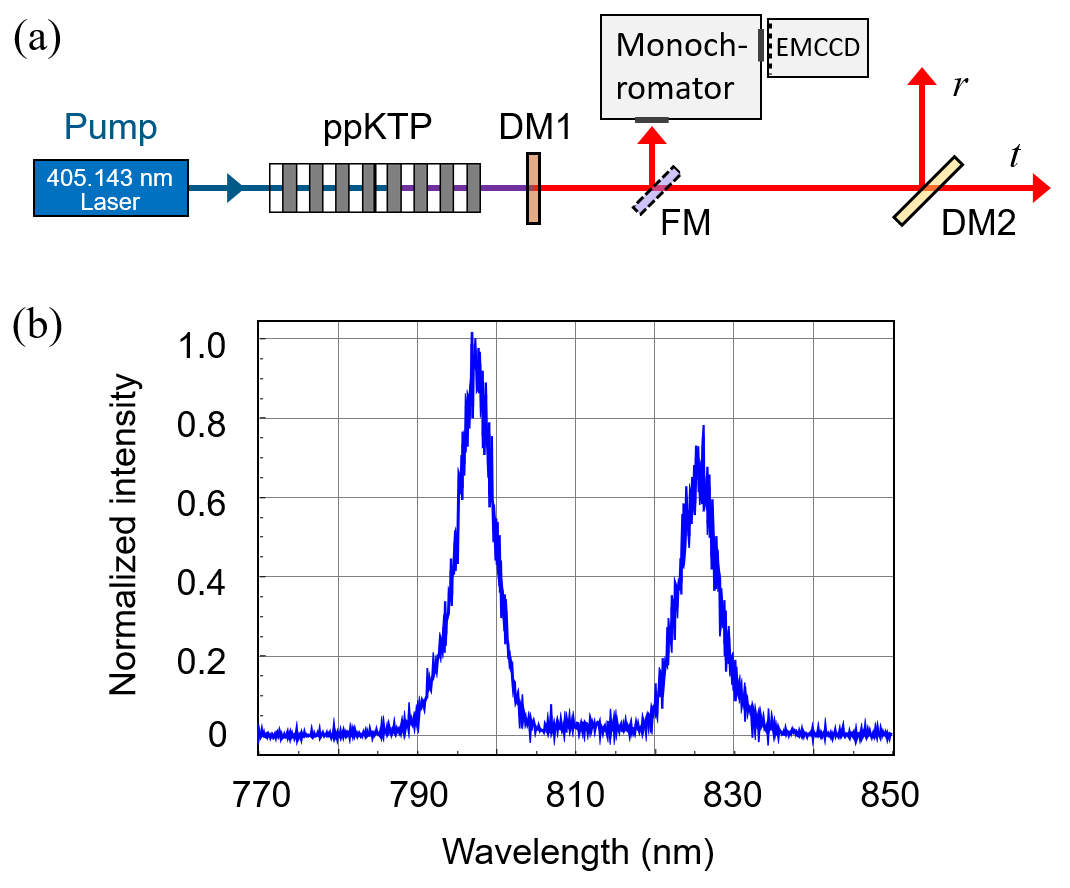}
                \caption{(a) Experimental setup for photon source where the ppKTP crystal satisfying the degenerate down-conversion of 405 nm pump photon to 810 nm photon pair is pumped by a ultra-narrowband pump laser operating at 405.143 nm center wavelength. DM, dichroic mirror; FM, flip mirror. (b) Spectrum for signal and idler photons measured by a monochromator and EMCCD. The left spectral peak is centered at 797.3 nm with a full width at half maximum (FWHM) of 5.77 nm, while the right peak is centered at 825.4 nm with a FWHM of 5.97 nm.}
                \label{fig01}
            \end{figure}

The experimental setup for photon pair generation is schematically shown in Fig.~\ref{fig01}(a). A continuous-wave diode laser (Topmode 405HP, Toptica) operating at 405.143 nm with a linewidth below 0.01 pm pumps a 30-mm-long periodic-poled potassium titanyl phosphate (ppKTP) crystal under collinear type-0 phase-matching conditions. The poling period $\Lambda$ is 3.425 $\mu$m which satisfies the degenerate down-conversion of 405 nm pump photon to 810 nm photon pair and the crystal temperature is stabilized at 29.3$^\circ$C. The pump beam first progagates through the crystal, then photon pairs are generated collinearly and co-polarized. After the crystal, the residual pump is removed by a long-pass dichroic mirror (DM1) with a cutoff wavelength of 650 nm. The generated photon pair is then measured using a monochromator and an electron-multiplying charge-coupled device (EMCCD) after reflected by a flip mirror (FM) and the result is shown in Fig.~\ref{fig01}(b). Unlike conventional type-0 SPDC pumped with a few nm bandwidth pump laser, which produces a smooth and broad spectrum centered at degeneracy, our source exhibits a clear spectrally split structure with a suppressed central region. The left spectral peak is centered at 797.3 nm with a full width at half maximum (FWHM) of 5.77 nm, while the right peak is centered at 825.4 nm with a FWHM of 5.97 nm. This behavior originates from a detuning of the pump center wavelength from 405 nm, combined with an ultra-narrow pump bandwidth that does not cover the degenerate phase-matching condition. Importantly, this spectral feature does not imply that one peak corresponds to the signal photon and the other to the idler photon. Instead, both photons share the same spectral distribution, and each photon individually exhibits a double-peaked spectrum.

To characterize the performance of the generated photon source, the photon pairs are separated using DM2 with a cutoff wavelength of 805 nm as shown in Fig.~\ref{fig01}(a), and the reflected and transmitted photons are coupled into single-mode fibers and detected using single photon detectors. Under a pump power of 15 $\mu$W, we measured a coincidence rate of 44.6 kHz between the two output modes. As the reflected and transmitted photons exhibit spectral bandwidths of 5.77 nm and 5.97 nm, we use their marginal bandwidth, $\Delta \lambda_{bp}$ = 5.87 nm, as an estimate of the biphoton spectral bandwidth. Then, we obtain a detected coincidence spectral brightness of 0.507 MHz mW$^{-1}$nm$^{-1}$. Here, the spectral brightness is calculated directly from the measured coincidence rate without correcting for optical transmission losses, fiber-coupling losses, or detector efficiencies. The result confirms that the spectrally engineered type-0 source maintains a high photon-pair flux within a narrow spectral bandwidth.

\begin{figure}[t!]
                \centering
                \includegraphics[width=0.53\textwidth]{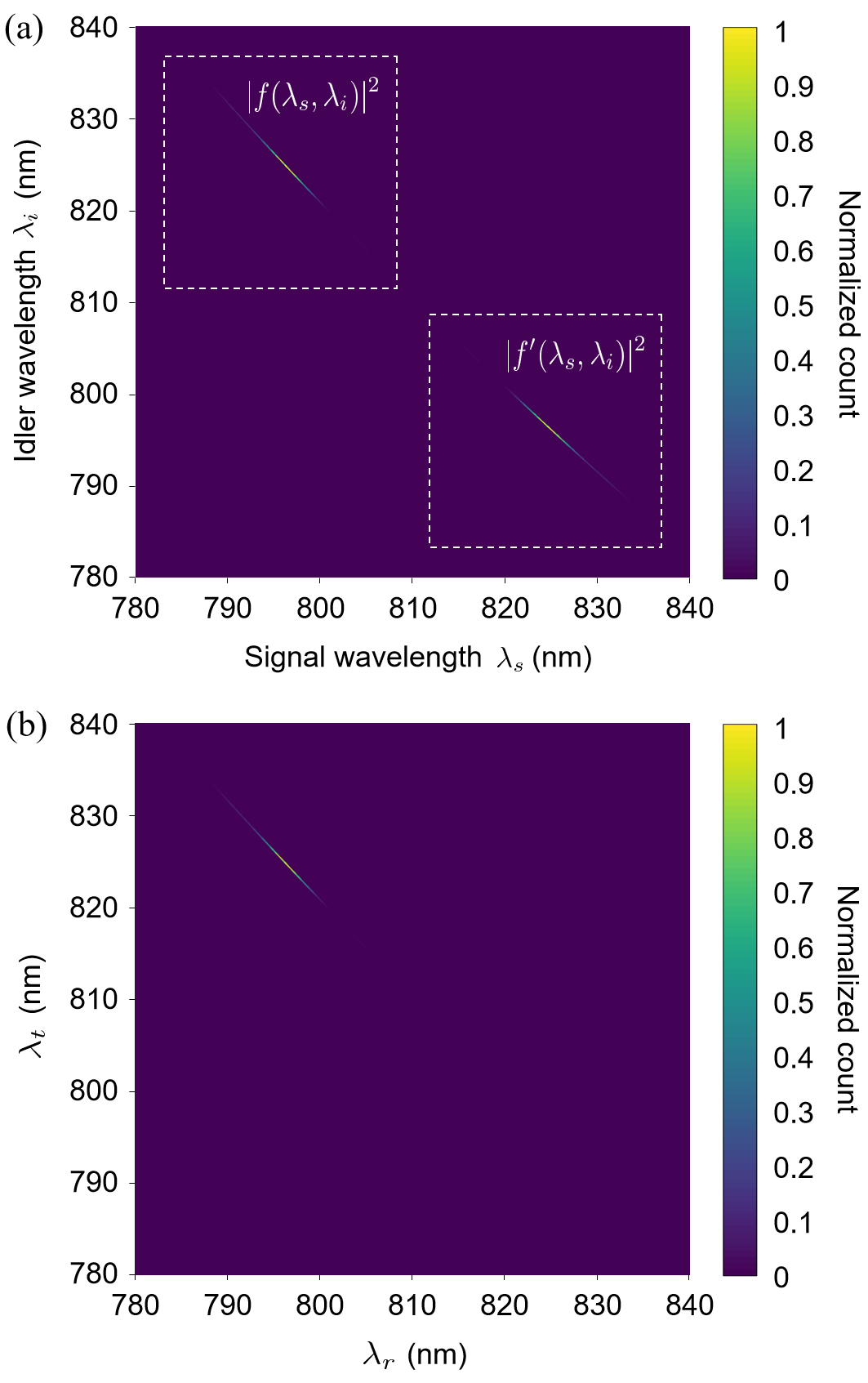}
                \caption{(a) JSI before passing through DM2 shown in Fig~\ref{fig01}(a). The result clearly shows the spectrally split structure. (b) JSI after passing through DM2. The reflected and transmitted photons show a single continuous joint spectrum. }
                \label{fig02}
            \end{figure}

To quantitatively simulate the observed spectrum, we calculate the joint spectral intensity (JSI). The quantum state of SPDC is written as,
	\begin{align} \label{eq01}
|\Psi\rangle = \int d\omega_s d\omega_i F(\omega_s,\omega_i) \hat{a}_s^\dagger(\omega_s) \hat{a}_i^\dagger (\omega_i) |0\rangle,
	\end{align}
where $\omega_{s,i}$ and $\hat{a}_{s,i}^\dagger(\omega_{s,i})$ denote the frequency and the creation operator for the signal and idler, respectively.
The joint spectral amplitude (JSA) $F(\omega_s,\omega_i)$ is defined as the product of a spectral distribution function of the pump and a phase-matching function, $F(\omega_s,\omega_i) = S_p(\omega_s,\omega_i) \Phi(\omega_s, \omega_i)$. The spectral distribution of the pump is given by $S_p(\omega_s,\omega_i) = \text{Exp}[-((\omega_p-(\omega_s+\omega_i))/\sigma_p)^2]$ where $\omega_p$ and $\sigma_p$ are the center frequency and bandwidth of pump, and the phase mismatching term is expressed as $\Phi(\omega_s, \omega_i) = \text{sinc}(\Delta k(\omega_s, \omega_i) L/2)$ where $L$ is the crystal length and the phase mismatch is given by $\Delta k(\omega_s, \omega_i) = k_p(\omega_s+\omega_i) - k_s(\omega_s) - k_i(\omega_i) - 2\pi/\Lambda$. Figure \ref{fig02}(a) shows the calculated JSI, $|F(\omega_s,\omega_i)|^2$ in the wavelength basis under the experimental conditions, reproducing the spectrally split structure observed experimentally.
As the pump center wavelength is detuned from the exact degeneracy condition at 405 nm, the product of the pump spectral function and the phase-mismatching function overlaps only in two distinct spectral regions. 

Note that for collinear type-0 phase matching in ppKTP, the pump, signal and idler fields propagate collinearly and share the same polarization. Consequently, the signal and idler photons experience identical dispersion relation, $k_s(\omega) = k_i(\omega) = k(\omega)$, and the phase mismatch reduces to $\Delta k(\omega_s, \omega_i) = k_p(\omega_s+\omega_i) - k(\omega_s) - k(\omega_i) - 2\pi/\Lambda$. Therefore, the phase-matching function satisfies $\Phi(\omega_s, \omega_i) = \Phi(\omega_i, \omega_s)$. 
 In addition, $S_p(\omega_s,\omega_i)$ which depends only on the sum frequency $\omega_s+\omega_i$ satisfies $S_p(\omega_s,\omega_i) = S_p(\omega_i,\omega_s)$. As a result, the JSA $F(\omega_s,\omega_i)$ exhibits exchange symmetry with respect to the siganl and idler frequencies for collinear type-0 SPDC in ppKTP.

Intriguingly, the frequency-exchange symmetry enables deterministic spatial separation of photon pair with a fixed joint spectrum using a single DM.
As discussed above, the JSA of the SPDC source employed in our experiment exhibits a spectrally split structure. We therefore decompose the $F(\omega_s,\omega_i)$ into two components, denoted by $f(\omega_s,\omega_i)$ and $f'(\omega_s,\omega_i)$, corresponding to the two spectrally separate regions, as indicated by the dashed rectangles in Fig.~\ref{fig02}(a). Note that Fig.~\ref{fig02}(a) shows the JSI in the wavelength basis, and the two components are indicated by the square of the JSA. Then, we can rewrite the quantum state as, 
	\begin{equation}
		\begin{aligned} \label{eq02}
|\Psi\rangle = \int d\omega_s d\omega_i (f(\omega_s,\omega_i) + f'(\omega_s,\omega_i)) \hat{a}_s^\dagger(\omega_s) \hat{a}_i^\dagger (\omega_i) |0\rangle.
		\end{aligned}
	\end{equation}
After passing through the long-pass DM2 with a cutoff wavelength of 805 nm shown in Fig.~\ref{fig01}(a), the quantum state becomes,
	\begin{equation}
		\begin{aligned} \label{eq03}
 \rightarrow \int d\omega_r d\omega_t &\big( f(\omega_r,\omega_t) \hat{a}_r^\dagger(\omega_r) \hat{a}_t^\dagger (\omega_t) 
 \\& + f'(\omega_t,\omega_r) \hat{a}_t^\dagger(\omega_t) \hat{a}_r^\dagger (\omega_r) \big) |0\rangle,
		\end{aligned}
	\end{equation}
where $\hat{a}^\dagger_r \,(\hat{a}^\dagger_t)$ is a creation operator for the reflected (transmitted) path mode shown in Fig.~\ref{fig01}(a).
As the JSA $F(\omega_s,\omega_i)$ exhibits frequency-exchange symmetry, $f'(\omega_t,\omega_r)$ is equal to $f(\omega_r,\omega_t)$, then the quantum state becomes,
\begin{align} \label{eq04}
|\Psi\rangle_{DM2} = \int d\omega_r d\omega_t f(\omega_r,\omega_t) \hat{a}_r^\dagger(\omega_r) \hat{a}_t^\dagger (\omega_t) |0\rangle.
	\end{align}
Here, $f(\omega_r,\omega_t)$ is a JSA for the reflected and transmitted photons, and Figure \ref{fig02}(b) shows the JSI, $|f(\omega_r,\omega_t)|^2$ in the wavelength basis.
Two processes contribute in this configuration: the signal photon is reflected at DM2 while the idler photon is transmitted, or vice versa. Since these two processes yield identical JSIs for the reflected mode $r$ and transmitted mode $t$, they are indistinguishable, resulting in deterministic spatial separation of the photon pair while preserving a fixed joint spectral distribution.

\subsection{Entanglement distribution via 2.56 km free-space}
\begin{figure}[t!]
                \centering
                \includegraphics[width=0.52\textwidth]{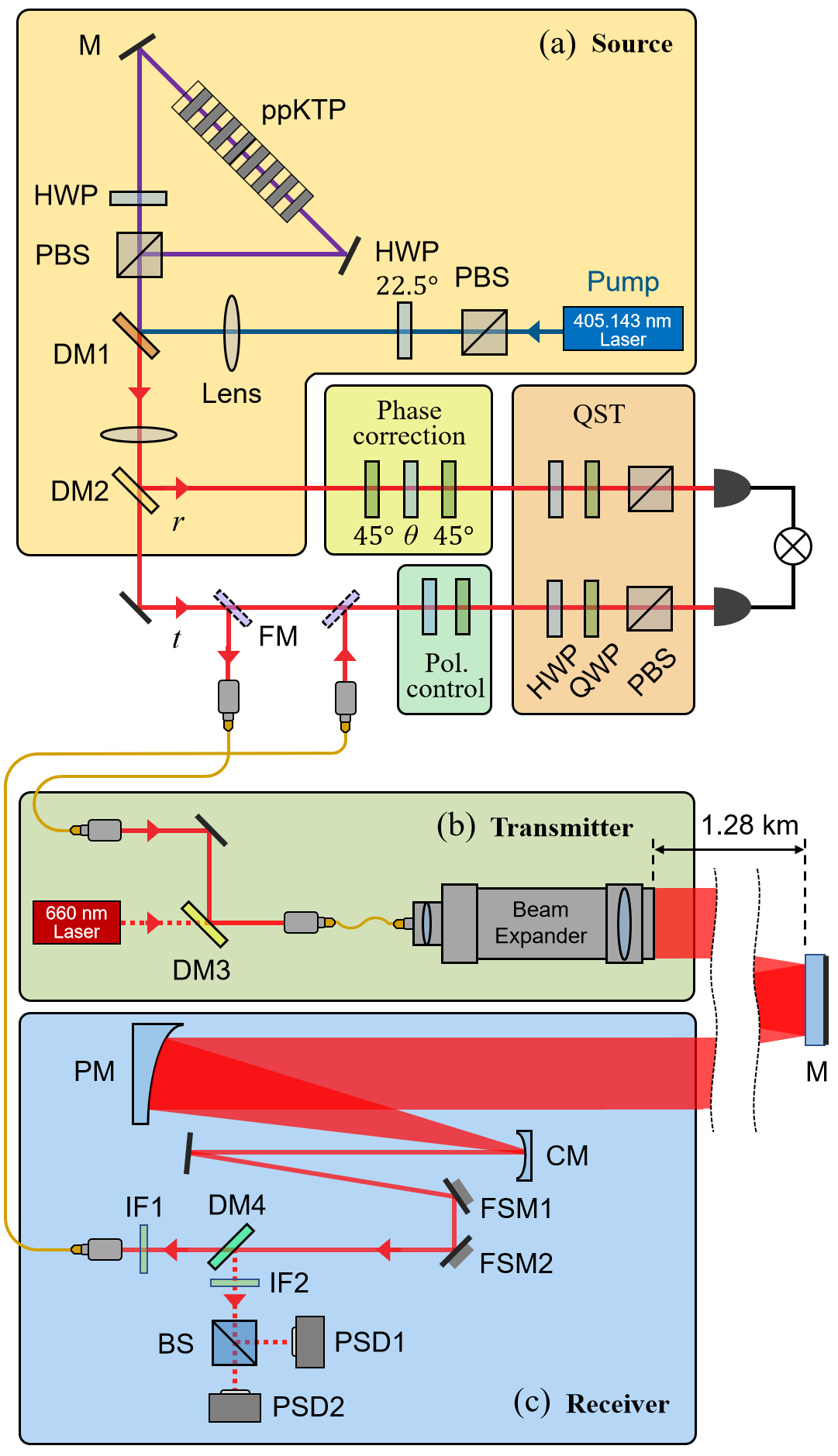}
                \caption{Experimental setup for polarization entanglement source and entanglement distribution via 2.56 km free-space. (a) Polarization entanglement source using a Sagnc interferometer with the spectrally engineered collinear type-0 SPDC source. (b) and (c) are the transmitter and receiver system, respectively. M, mirror; HWP, half-wave plate; QWP, quarter-wave plate; DM, dichroic mirror; FM, flip mirror; PBS, polarizing beam splitter; BS, beam splitter; PM, off-axis parabolic mirror; CM, concave mirror; FSM, fast steering mirror; PSD, position-sensitive detector; IF, interference filter}
                \label{fig03}
            \end{figure}

By making use of our spectrally engineered photon source, we generate polarization entanglement in a Sagnac interferometer as shown in Fig.~\ref{fig03}(a).
The pump laser operating at 405.143 nm passes through a polarizing beam splitter (PBS) and an half-wave plate (HWP) to become diagonally polarized and goes into the Sagnac interferometer after reflected at DM1.
Then, the pump laser is divided into two pathways by PBS and propagates along the clockwise and counterclockwise pathways to pump a 30 mm-long ppKTP crystal with a 3.425 um poling period. 
 The clockwise-propagating pump generates photon pairs in the $|VV\rangle$ state, whereas the counterclockwise-propagating pump generates photon pairs in the $|HH\rangle$ state at the output of the Sagnac interferometer. Finally, the coherent superposition of these two processes gives rise to the polarization entanglement $\frac{1}{\sqrt2}(|HH\rangle +e^{i \phi}|VV\rangle)$ where $\phi$ is a relative phase between $|HH\rangle$ and $|VV\rangle$. 

Before the spectral separation at DM2, the polarization entangled signal and idler photons has an JSI shown in Fig.~\ref{fig02}(a) and the quantum state is described as,
	\begin{equation}
		\begin{aligned} \label{eq05}
|\Psi\rangle = \int d\omega_s d\omega_i F(\omega_s,\omega_i) &\big( \hat{a}_{s,H}^\dagger(\omega_s) \hat{a}_{i,H}^\dagger (\omega_i) 
\\& + \hat{a}_{s,V}^\dagger(\omega_s) \hat{a}_{i,V}^\dagger (\omega_i) \big) |0\rangle.
		\end{aligned}
	\end{equation}
However, after the spectral separation at DM2, the two photons are spatially separated to the reflected and transmitted modes deterministically which has the JSI shown in Fig.~\ref{fig02}(b) keeping the polarization entanglement as, 
	\begin{equation}
		\begin{aligned} \label{eq06}
|\Psi\rangle = \int d\omega_r d\omega_t f(\omega_r,\omega_t) &\big( \hat{a}_{r,H}^\dagger(\omega_r) \hat{a}_{t,H}^\dagger (\omega_t) 
\\& + \hat{a}_{r,V}^\dagger(\omega_r) \hat{a}_{t,V}^\dagger (\omega_t) \big) |0\rangle.
		\end{aligned}
	\end{equation}
Note that the two photons are no longer labeled as signal and idler but are instead referred to as the reflected ($r$) and transmitted ($t$) photons. 

The $r$ photon propagates through a QWP(45$^\circ$)-HWP-QWP(45$^\circ$) waveplate sequence, which compensates the relative phase $\phi$, and is then directed to quantum state tomography (QST) setup followed by a single-photon detector.
The $t$ photon is coupled into a single-mode fiber and delivered to the transmitter module, where the $t$ photon is combined with a 660 nm classical beam at DM3 as shown in Fig.~\ref{fig03}(b). The combined beams are subsequently expanded by a beam expander ($ 20\times$) and launched into free-space. The expanded beam propagating over a 1.28 km free-space, is reflected by a distant mirror and returns to the laboratory forming a 2.56 km round-trip channel.
In the receiver represented in Fig.~\ref{fig03}(c), the returning beam sequentially passes through a off-axis parabolic mirror (PM) and a concave mirror (CM), which are used to reduce the beam size and adapt it to the receiver optics, and is then reflected by two fast steering mirrors (FSMs). At DM4, the $t$ photon is transmitted while the 660 nm classical light is reflected. The reflected classical beam is detected by two position-sensitive detectors (PSDs), and the extracted positional information is fed back to FSM1 and FSM2 to actively compensate beam wandering induced by an atmospheric turbulence \cite{Lim23}. Finally, the stabiliezd $t$ photon is coupled back into a single-mode fiber, routed to the measurement setup after polarization compensation.

\begin{figure}[!t]
                \centering
                \includegraphics[width=0.52\textwidth]{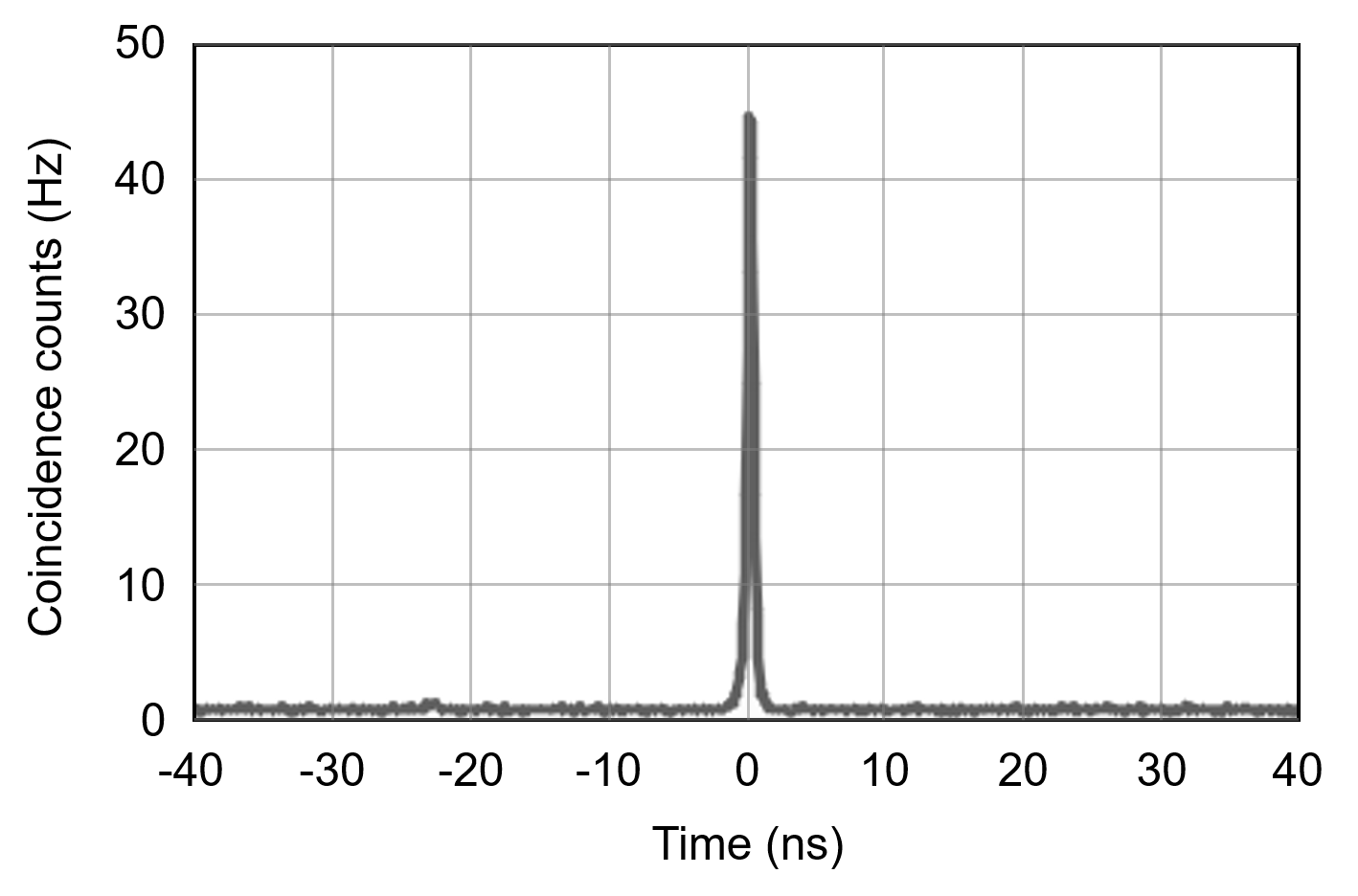}
                \caption{Temporal correlation between the locally detected $r$ photon and the $t$ photon transmitted through the 2.56 km free-space channel, measured using a TCSPC system. After compensating the propagation delay, a coincidence peak appears at zero relative delay.}
                \label{fig04}
            \end{figure}

To first verify that photon pair correlations are preserved after transmission, we measured the temporal correlation between the locally detected $r$ photon and the $t$ photon transmitted through the free-space link and subsequently received in the laboratory using a time-correlated single-photon counting (TCSPC) system. Since the $t$ photon propagates over the round-trip free-space path, the arrival time difference between the two photons corresponds to the channel propagation delay. By applying an electronic delay of 8.537700 $\mu$s to the locally detected $r$ photon, a coincidence peak appears at zero relative delay, as shown in Fig.~\ref{fig04}, obtained during nighttime free-space transmission. This confirms that photon pairs retain temporal correlation after free-space transmission and reception, demonstrating successful pairwise distribution through the atmospheric channel. After free-space transmission and reception at nighttime, the detected coincidence rate was measured to be 294 Hz, which is consistent with the temporal correlation measurement shown in Fig.~\ref{fig04} when a coincidence window of 1 ns is considered. 
The reduction in coincidence counts is mainly attributed to geometric propagation losses, imperfect optical alignment, and coupling losses at the receiver. Furthermore, atmospheric turbulence along the free-space channel introduces wavefront distortion, resulting in additional fluctuations in the coupling efficiency into the single-mode fibers.

\begin{figure}[t!]
                \centering
                \includegraphics[width=0.55\textwidth]{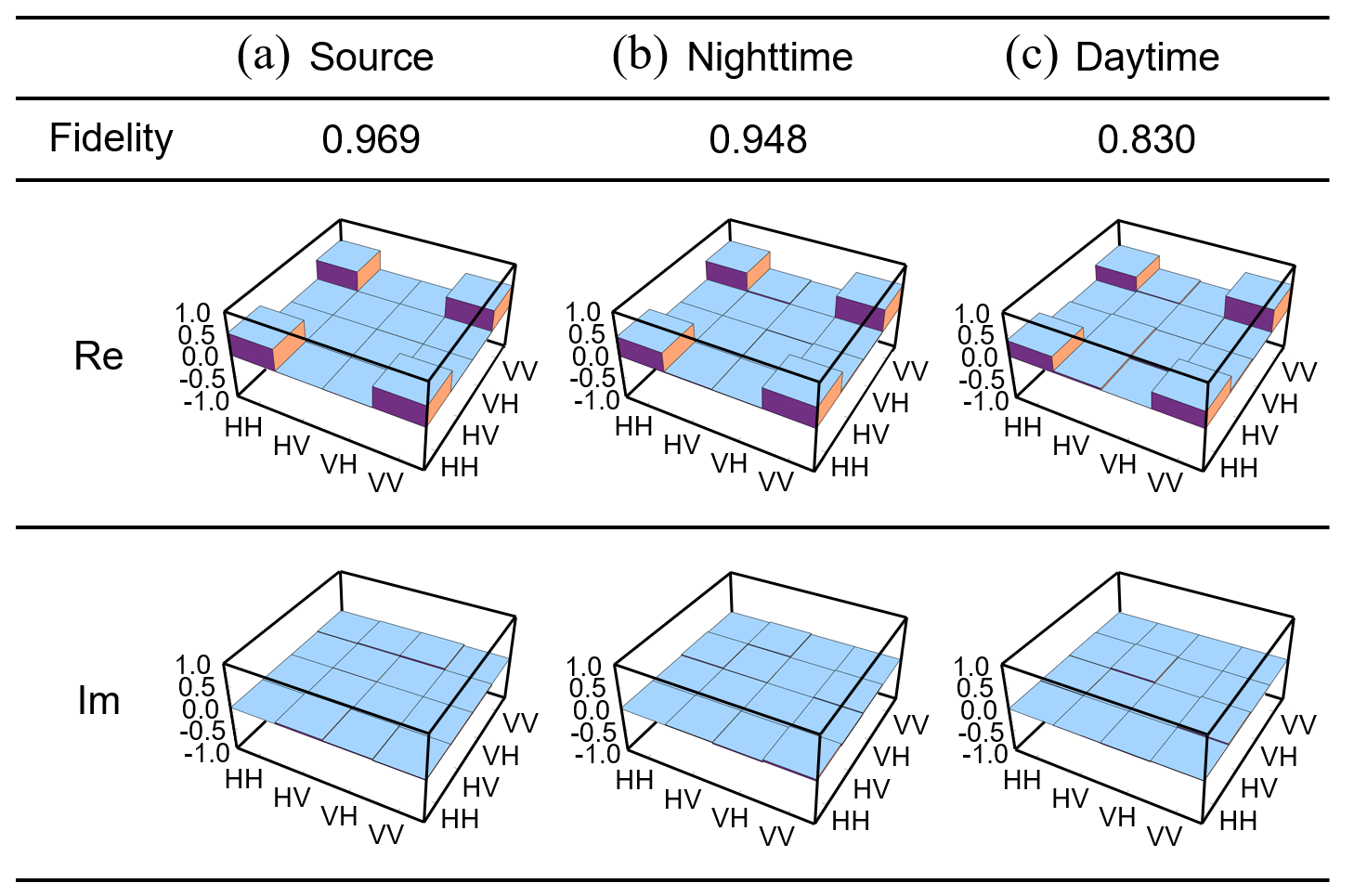}
                \caption{Results for polarization entanglement distribution. Density matrices reconstructed via QST and the corresponding entanglement fidelities are shown for (a) the local reference, (b) the nighttime free-space transmission, and (c) the daytime free-space transmission.}
                \label{fig05}
            \end{figure}

Following this verification, we performed QST to characterize the polarization state of the distributed photon pairs. Figure \ref{fig05} summarizes the density matrices reconstructed via QST and the corresponding entanglement fidelities. The first column shows the results for the local reference measurements confirming polarization entanglement at the source, while the second and third columns correspond to the results obtained after nighttime and daytime free-space transmission of the $t$ photon, respectively. The nighttime results are consistent with the local reference, as backgound noise from sunlight is negligible. The slight degradation observed compared to the local reference is mainly attributed to polarization and phase drifts in the optical fiber during the measurement. In contrast, the daytime results exhibit reduced fidelity due to increased background noise from sunlight. Although a bandpass filter was employed to suppress the background light, its 40 nm full width at half maximum (FWHM) bandwidth was insufficient to suppress the solar noise. These results indicate that the observed degradation primarily arises from insufficient background suppression rather than from the intrinsic performance of the entanglement source.

\section{Conclusion}
In conclusion, we demonstrated a spectrally engineered collinear type-0 SPDC source with enhanced detected spectral brightness. By employing an ultra-narrowband pump laser slightly detuned from the degenerate phase-matching condition, photon-pair generation was concentrated into a narrow spectral region, resulting in a spectrally split emission. Under a pump power of 15 $\mu$W, the source produced a coincidence rate of 44.6 kHz, corresponding to a detected conicidence spectral brightness of 0.507 MHz mW$^{-1}$nm$^{-1}$. By incorporating the source into a Sagnac interferometer, we realized efficient polarization-entangled photon generation and successfully demonstrated free-space entanglement distribution over a 2.56 km round-trip channel, where both temporal correlations and polarization entanglement were preserved after transmission

By increasing the usable photon-pair flux within a narrow bandwidth, the present approach improves the source-level efficiency of entanglement generation without relying on post-generation spectral filtering. This feature is particularly relevant for quantum networks, where channel loss and source inefficiency limit entanglement distribution rates and scalability. Spectrally engineered type-0 SPDC therefore provides a practical route toward compact, efficient, and deployable entangled-photon sources for free-space quantum communication and quantum network.

\begin{backmatter}
\bmsection{Funding}
This work was supported by the Agency for Defense Development Grant funded by the Korean Governmnet.

\bmsection{Disclosures}
The authors declare no conflicts of interest.

\bmsection{Data availability} Data underlying the results presented in this paper are not publicly available at this time but may be obtained from the authors upon reasonable request.
\end{backmatter}


\end{document}